\documentclass[epj]{svjour}
\usepackage{graphics}
\usepackage{epsfig}
\def\be{\begin{equation}}
\def\ee{\end{equation}}
\def\bq{\begin{eqnarray}}
\def\eq{\end{eqnarray}}
\def\R{\mbox{${\cal R}$}}
\begin{document}

\title{Quark matter nucleation in neutron stars and astrophysical implications}
\author{Ignazio Bombaci\inst{1,2,3}, Domenico Logoteta \inst{2}, Isaac Vida\~na \inst{4} \and 
 Constan\c{c}a Provid\^encia \inst{4} 
}                     
\offprints{ignazio.bombaci@unipi.it} 
\institute{Dipartimento di Fisica ``E. Fermi'', Universit\`{a} di Pisa, Largo B. Pontecorvo, 3, I-56127 Pisa, Italy  
\and  
INFN, Sezione di Pisa, Largo B. Pontecorvo, 3, I-56127 Pisa, Italy 
\and 
European Gravitational Observatory, Via E. Amaldi, I-56021 S. Stefano a Macerata, Cascina Italy  
\and 
CFisUC, Department of Physics, University of Coimbra, PT-3004-516 Coimbra, Portugal}

\date{Received: date / Revised version: date}

\abstract{A phase of strong interacting matter with deconfined quarks is expected in 
the core of massive neutron stars. We investigate the quark deconfinement phase transition 
in cold ($T = 0$) and hot $\beta$-stable hadronic matter. 
Assuming a first order phase transition, we calculate and compare the nucleation rate and the 
nucleation time due to quantum and thermal nucleation mechanisms.  
We show that above a threshold value of the central pressure a pure hadronic star (HS)  
({\it i.e.} a compact star with no fraction of deconfined quark matter) is metastable to the 
conversion to a quark star (QS) ({\it i.e.} a hybrid  star or a strange star).  
This process liberates an enormous amount of energy, of the order of 10$^{53}$~erg, 
which causes a powerful neutrino burst, likely accompanied by intense gravitational waves emission, 
and possibly by a second delayed (with respect to the supernova explosion forming the HS) 
explosion which could be the energy source of a powerful gamma-ray burst (GRB).  
This stellar conversion process populates the QS branch of compact stars, thus one has in the 
Universe two coexisting families of compact stars: pure hadronic stars and quark stars. 
We introduce the concept of critical mass $M_{cr}$ for cold HSs and proto-hadronic stars (PHSs),   
and the concept of limiting conversion temperature for PHSs.   
We show that PHSs with a mass $M < M_{cr}$ could survive the early stages of their evolution  
without decaying to QSs. Finally, we discuss the possible evolutionary paths of proto-hadronic stars.
} 

\authorrunning{Ignazio Bombaci {\it et al.}}
\titlerunning{Quark matter nucleation in neutron stars and astrophysical implications}
\maketitle

\section{Introduction}
\label{intro}
Neutron stars, the compact remnants of core-collapse supernova, are the densest 
macroscopic objects in the Universe. They represent the limit beyond which gravity 
overwhelm all the other forces of nature and lead to the formation of a black hole.  
In fact, neutron star structure calculations 
(see {\it e.g.} Refs.~\cite{lp01,peng08,li08}) based on a large variety of 
modern equations of state (EOS) of hadronic matter,  predict a maximum stellar central density  
(the one for the maximum  mass star configuration) in the range of  4 -- 8 times the saturation 
density ($\sim 2.8 \times 10^{14}$~g/cm$^{3}$) of nuclear matter.  
Thus the core of a neutron star is one of the best candidates in the Universe where a 
phase of strong interacting matter with deconfined quarks could be  found, 
and these compact stars can be viewed as natural laboratories to test the low temperature 
$T$ and high baryon chemical potential $\mu$ region of the QCD phase diagram. 

 Current high precision numerical calculations of QCD on a space-time lattice at zero baryon chemical 
potential (zero baryon density) have shown that at high temperature 
and for physical values of the quark masses, the transition to quark gluon plasma is a 
crossover \cite{bern05,cheng06,aoki06} rather than a real phase transition.  

Unfortunately, present lattice QCD calculations at finite baryon chemical potential are 
plagued with the so called ``sign problem'', which makes them unrealizable by all 
presently known lattice methods.      
Thus, to explore the QCD phase diagram at low T and high $\mu$, it is 
necessary to invoke some approximations in QCD or to apply a QCD effective model 
\cite{njl,bub05,meisinger96,fukushima04,ratti06,blaschke08,contrera08,blaschke10,dexh10,gie12}. 
In this region of the T-$\mu$ plane, several QCD inspired models suggest the deconfinement transition 
to be a first-order phase transition \cite{hs98,fk04}. 
In this domain of the QCD phase diagram, many possible color superconducting phases of quark matter  
are expected \cite{CN04,alf+08} and matter might be characterized by the formation of 
different crystalline structures \cite{angl14,bub15}.  
It is worth mentioning that recent promising attempts to describe the whole QCD phase diagram 
within a unified model \cite{st,digiac02,ST07,sim5} (see also \cite{david15}) 
provides a powerful tool to link numerical lattice QCD calculations with measured neutron 
star masses \cite{BL13,LB13,bz15}.  
 
Here, we have adopted a more traditional and simple view, assuming 
a single first-order phase transition between the confined (hadronic) and deconfined phase 
of dense matter, and we  used rather common models for describing them. 

As it is well known, all first order phase transitions are triggered by the nucleation of a  
critical size drop of the new (stable) phase in a metastable mother phase. 
This is a very common phenomenon in nature (e.g. fog or dew formation in supersaturated vapor, 
ice formation in supercooled water) and plays an important role in many scientific disciplines 
(e.g. atmospheric science, meteorology, cosmology, biology) as well as in 
many technical applications (e.g. metallurgy). 
  
In the last few years, we have investigated 
\cite{be02,be03,bo04,vbp05a,vbp05b,lug05,blv07,bppv08,bom+09,bom10,bom+11,log+12a,log+12b,log+13}
(see also Refs.~\cite{drago04,dps-b08,bambi08,DLP07,min+10,lug+11,doCar+13,lug15}) 
the astrophysical consequences of the nucleation process of quark matter (QM) in the core 
of massive pure hadronic compact stars (hadronic stars, HSs)  {\it i.e.} neutron stars in 
which no fraction of QM is present.  
In this contribution, we report some of the main findings of these studies.  

\section{Equation of state of dense matter}
All the results we report in the present paper are relative to the zero and finite temperature 
version of the following models for the EOS of dense matter.
For the hadronic phase we use the Glendenning--Moszkowski model \cite{gm91,glen00}, 
and particularly the GM1 and the GM3 parametrizations \cite{gm91,glen00,bppv08}.  
The nucleon coupling constants are fitted to the bulk properties of nuclear matter.  
The inclusion of hyperons involves  new couplings, which can be written in terms of 
the nucleonic ones as: 
  $g_{\sigma Y}=x_{\sigma }~ g_{\sigma},~~g_{\omega Y}  =x_{\omega }~ g_{\omega},
  ~~g_{\rho Y}=x_{\rho }~ g_{\rho}$. 
In this model \cite{gm91,glen00} it is assumed that all the hyperons in the baryonic octet 
have the same coupling and, in addition, it is assumed that $x_{\rho} = x_{\sigma}$. 
The binding energy of the $\Lambda$ particle in symmetric nuclear matter   
$ B_\Lambda/A  = - 28 \mbox{ MeV} = 
x_{\omega} \, g_{\omega}\, \omega_0 - x_{\sigma}\,  g_{\sigma} \sigma \,\,$
is used \cite{gm91,glen00} to determine $x_\omega$ in terms of $x_\sigma$.    
In this work we will consider the cases $x_\sigma=0.6$ (hereafter GM1$_{0.6}$)  
$x_\sigma=0.7$ (GM1$_{0.7}$),  and $x_\sigma=0.8$ (GM1$_{0.8}$).    
Notice that the case with $x_{\sigma} = 0.6$ produces stars with a larger hyperon population 
(for a given stellar gravitational mass) with respect to the case  $x_{\sigma} = 0.7$ and 
$x_{\sigma} = 0.8$ \cite{glen00,bppv08}. 

For the deconfined quark phase we have used the following models:  
(i) the MIT bag model EOS \cite{fj84} with $m_s = 150~\rm{MeV}$, $m_u = m_d=0$, $\alpha_s = 0$ and different values for the bag constant $B$;  
(ii) an extended version of the MIT bag model EOS which includes  perturbative 
corrections due to quark interactions, up to the second order ($\mathcal{O}(\alpha_{\rm s}^2)$) in 
the strong structure constant $\alpha_s$  \cite{Fra01,Alf05,weis11}.  
This EOS model is parametrized in term of an effective bag constant ($B_{\rm eff}$) and 
a perturbative QCD correction term ($a_4$), whose value represents the degree of deviations 
from an ideal relativistic Fermi gas EOS, with the case $a_4 = 1$ corresponding to the ideal 
gas \cite{Fra01,Alf05,weis11}. Within this extended bag model one can thus evaluate the 
non-ideal behaviour of the EOS of cold SQM at high density; 
(iii) the Nambu--Jona-Lasinio (NJL) model \cite{njl}, with the lagrangian density given 
in Ref.~\cite{njl2}; 
(iv) the Chromo Dielectric model \cite{cdm,cdm9}. 

\section{Phase equilibrium}
For a first-order phase transition the conditions for phase equilibrium  are given by 
the Gibbs' phase rule
\begin{eqnarray}
T_H  = T_Q \equiv T   \, , ~~~~~~~~ P_H = P_Q \equiv P_0 \, , \nonumber \\ 
\mu_H(T, P_0)  =  \mu_Q(T, P_0) \, 
\label{eq:eq1bis}
\end{eqnarray}
where 
\begin{equation}
  \mu_H = \frac{\varepsilon_H + P_H - s_H T}{n_H}  \, ,~~~~~~~~
  \mu_Q = \frac{\varepsilon_Q + P_Q - s_Q T}{n_Q}  
\label{eq:eq2}
\end {equation} 
are the Gibbs energies per baryon (average chemical potentials) for the hadron and 
quark phase respectively, 
$\varepsilon_H$ ($\varepsilon_Q$),  $P_H$ ($P_Q$), $s_H$ ($s_Q$)  and $n_{H}$  ($ n_{Q}$)
denote respectively the total ({\it i.e.,}  including leptonic contributions) energy 
density, total pressure, total entropy density,  and baryon number density  for the hadron (quark)  
phase.  

\begin{figure}
\vspace{0.5cm}
\includegraphics[scale=0.33, angle = 270 ]{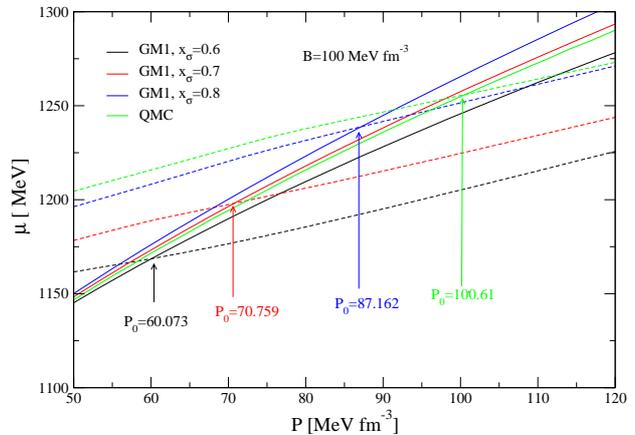} 
\caption{(Color on line) The Gibbs energy per particle for the $\beta$-stable 
hadronic phase (continuous curves) and for the respective $Q^*$ phase (dashed curves) at $T = 0$. 
The GM1 and the QMC models have been used for the hadronic phase EOS, 
and bag model EOS with B = 100 MeV/fm$^3$ for the Q*-phase.} 
\label{gibbs_T0}
\end{figure}

Above the transition point $P_0$ (see Fig.\ \ref{gibbs_T0}) the hadronic phase is metastable, 
and the stable quark phase will appear  as a result of a nucleation process. 
Virtual drops of the stable quark phase will arise from localized fluctuations in the state 
variables of the metastable hadronic phase. These fluctuations are characterized by a time scale  
$\nu_0^{-1} \sim 10^{-23}$ s. This time scale is set by the strong interactions (which are 
responsible for the deconfinement phase transition), and it is many orders of magnitude 
shorter than the typical time scale for the weak interactions.  
Therefore quark flavor must be conserved forming a virtual drop of QM. We will refer to this form 
of deconfined matter,  in which the flavor content is equal to that of the $\beta$-stable hadronic system at the same pressure and temperature, as the Q*-phase.  
For example, if quark deconfinement occurs in $\beta$-stable nuclear matter (non-strange hadronic matter), it will produce a two-flavor ($u$ and $d$ ) quark matter droplet having 
\begin{equation}
n_u / n_d = (1 + x_p)/(2-x_p) \,  ,     
\label{eq:eq3}
\end{equation}
$n_u$ and $n_d$ being the {\it up} and {\it down} quark number densities respectively, 
and $x_p$ the proton fraction in the $\beta$-stable hadronic phase.   
In the more general case in which the hadronic phase has a strangeness content 
({\it e.g.,} hyperonic matter), the deconfinement transition will form a droplet 
of strange matter with a flavor content equal to that of the $\beta$-stable 
hadronic system at the same pressure,  according to the relation:
\begin{equation}
\left( \begin{array}{c}
x_u \\ x_d \\ x_s
\end{array} \right)
=
\left( \begin{array}{cccccccc}
2 & 1 & 1 & 2 & 1 & 0 & 1 & 0 \\
1 & 2 & 1 & 0 & 1 & 2 & 0 & 1 \\
0 & 0 & 1 & 1 & 1 & 1 & 2 & 2
\end{array} \right)
\left( \begin{array}{c}
x_p \\ x_n \\ x_{\Lambda} \\ x_{\Sigma^+} \\
x_{\Sigma^0} \\ x_{\Sigma^-} \\ x_{\Xi^0} \\ x_{\Xi^-}
\end{array} \right) \, ,
\label{eq:eq4}
\end{equation}
where  $x_i = n_i/n$ are the concentrations of the different particle species.  

Soon afterward a critical size drop of Q*-matter is formed, the weak interactions 
will have enough time to act, changing the quark flavor fraction of the deconfined droplet to lower 
its energy, and a droplet of $\beta$-stable QM is formed (hereafter the Q-phase). 
This first seed of $\beta$-stable QM will trigger the conversion \cite{grb} of the pure hadronic 
star to a quark star (QS), {\it i.e.} to a hybrid neutron star or to a 
strange star \cite{bod71,witt84,afo86,hae86,dey98,li99a,li99b,Xu99} depending on the details of the EOS for 
quark matter used to model the phase transition.      
    
The direct formation by fluctuations of a drop of $\beta$-stable QM is also possible 
in principle. However, it is strongly suppressed with respect to the formation of the 
Q*-phase drop by a factor $\sim G_{\mathrm{F}}^{2N / 3}$, being $N$ the number 
of particles in the critical size quark drop and $G_F$ the Fermi constant of weak interaction. 
This is so because the formation of a $\beta$-stable drop will imply the almost simultaneous 
conversion of $\sim N/3$ up and down quarks into strange quarks.  
For a critical size $\beta$-stable drop at the center of a neutron star it is found 
$N \sim 100-1000$, and therefore the suppression factor is actually very tiny. 

\begin{figure}
\vspace{0.5cm}
\resizebox{0.43\textwidth}{!}{%
 \includegraphics{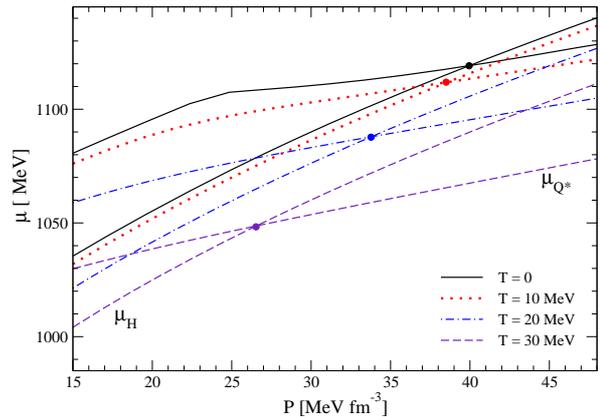}}
\caption{(Color on line) Gibbs energy per baryon of $\beta$-stable hadronic phase and Q*-phase,  
in neutrino-free matter, as a function of pressure, at different temperatures. 
Lines with the steeper slope refer to the hadronic phase. 
Full dots indicate the transition pressure $P_0$ for each temperature. 
GM1 EOS with $x_\sigma = 0.6$ for the hadronic phase and 
MIT bag model EOS with B = 85 MeV/fm$^3$ for the Q*-phase (GM1$_{0.6}$--B85 EOS). }
\label{gibbs_GM1_B85} 
\end{figure}

In Fig.\ \ref{gibbs_GM1_B85} we plot the Gibbs' energies per baryon for the hadron-phase 
and for the Q*-phase in neutrino-free matter, at different temperatures (T = 0, 10, 20, 30 MeV).  
Results in Fig.\ \ref{gibbs_GM1_B85} are obtained using the GM1 model with $x_\sigma = 0.6$ 
for the hadronic phase and the MIT bag model with B = 85 MeV/fm$^3$ for the quark phase 
(hereafter the GM1$_{0.6}$--B85 EOS).    
Lines with the steeper slope refer to the hadron phase.  
As we see, the transition pressure $P_0$ (indicated by a full dot) decreases when the 
hadronic matter temperature is increased.   

\begin{figure}
\vspace{0.5cm}
\resizebox{0.43\textwidth}{!}{%
 \includegraphics{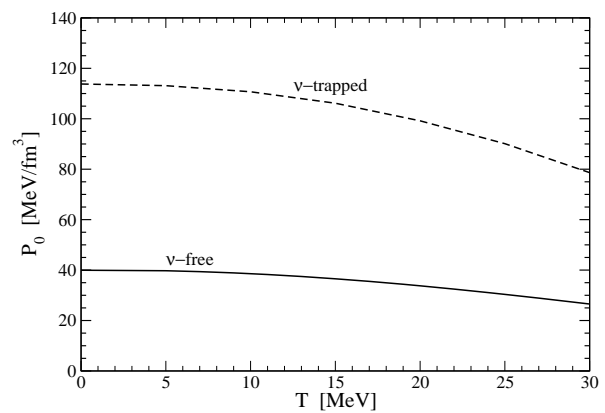}}
\caption{Phase equilibrium curve between the $\beta$-stable hadronic phase  and the  Q* phase.  
The continuous curve is relative to neutrino-free matter, 
the dashed curve to matter with trapped  neutrinos. 
EOS: GM1 with $x_\sigma = 0.6$  plus MIT bag model with B = 85 MeV/fm$^3$.}
\label{p0-T_GM1_B85}
\end{figure}

 The phase equilibrium curve $P_0(T)$ between the $\beta$-stable hadronic phase and 
the $Q^{*}$-phase is shown in  Fig.\ \ref{p0-T_GM1_B85} for neutrino-free matter and 
matter with trapped neutrinos, making use of the  GM1$_{0.6}$--B85 EOS. 
The region of the  $P_0$--$T$ plane above each curve represents the deconfined Q*-phase.  
As expected \cite{prak97,lb98,vbp05b,lugones09} neutrino trapping in $\beta$-stable 
hadronic matter inhibits the quark deconfinement phase transition, thus the global effect of 
neutrino-trapping is to produce a shift of the phase equilibrium curve toward higher values of 
the pressure in the  $P_0$--$T$ plane.   

\begin{figure}
\vspace{0.5cm}
\resizebox{0.43\textwidth}{!}{%
 \includegraphics{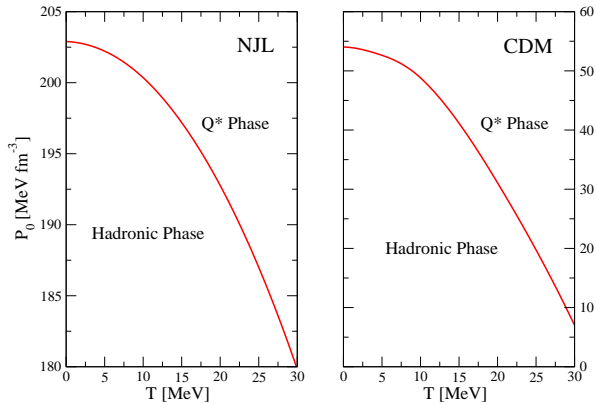}} 
\caption{Phase equilibrium curve between the $\beta$-stable hadronic phase and the Q*-phase, 
neutrino-free matter.   
The EOS for the hadronic phase is obtained using the  GM1 model with $x_{\sigma}=0.7$, 
the EOS for the $Q^{*}$-phase using the NJL model (left panel) 
or the Chromo Dielectric model (right panel).} 
\label{p0-T_GM1_NJL-CDM} 
\end{figure}

In Fig.\ \ref{p0-T_GM1_NJL-CDM}, we show the phase equilibrium curve, 
for neutrino-free matter, in the case of the NJL model (left panel)  
or the Chromo Dielectric model (right panel) to describe the deconfined phase. 
For the hadron phase we take the GM1 model with $x_\sigma = 0.7$  (GM1$_{0.7}$) 
in both cases.     
Notice that in the case of the NJL model the transition pressure $P_0(T)$ is substantially 
higher than the one in the case of the MIT bag or Chromo Dielectric models to describe the deconfined phase.  As discussed in detail in Ref. \cite{log+12a}, this behaviour can be traced back 
the large value of the strange quark effective mass in the NJL model.

As  is well known, for a first-order phase transition the derivative $dP_0/dT$ is related to the 
specific  latent heat ${\cal Q}$  of the phase transition by the  Clapeyron-Clausius equation   
\begin{equation}
            \frac{dP_0}{dT} =  - \frac{ n_H n_{Q^*}}{n_{Q^*} - n_H } \frac{{\cal Q}}{T } 
\label{eq:eq10}
\end {equation}
\begin{equation}
           {\cal Q}  = \tilde W_{Q^*} - \tilde W_H = T(\tilde {S}_{Q^*} - \tilde {S}_H) 
\label{eq:eq11}
\end {equation}
where $\tilde W_H$ ($\tilde W_{Q^*}$)  and $\tilde {S}_H$ ($\tilde {S}_{Q^*}$) denote the  enthalpy  
per baryon and entropy per baryon for the hadron (quark) phase, respectively.   
The specific latent heat ${\cal Q}$ and the hadron and quark baryon number densities $n_H$ and  $n_{Q^*}$ 
at phase equilibrium are reported in Tables 1 and 2 in the case of the GM1$_{0.6}$--B85  
equation of state.  Results in Table 1 refer to neutrino-free matter,  whereas 
those in Table 2 refer to matter with trapped neutrinos.  
As expected for a first-order phase transition, one has a discontinuity jump in the phase 
number densities: in our particular case  $n_{Q^*}(T,P_0) > n_H(T,P_0)$.   
This result, together with the positive value of ${\cal Q}$ ({\it i.e.} the deconfinement phase 
transition absorbs heat), tells us (see Eq.\ (\ref{eq:eq10})) that the phase transition 
temperature decreases with pressure (as in the melting of ice). 
   
The effect of neutrino trapping on the phase equilibrium properties of the system can be seen 
comparing the results reported in Tables 1 and 2.  As we see, the phase number  densities  
$n_H$ and  $n_{Q^*}$ at phase equilibrium are shifted to higher values, and the specific 
latent heat ${\cal Q}$  is increased with respect to the neutrino-free matter case.

\begin{table}
\caption{The specific ({\it i.e.} per baryon) latent heat ${\cal Q}$ and the phase number densities  
$n_H$ and  $n_{Q^*}$ at phase equilibrium. 
GM1 EOS with $x_{\sigma} = 0.6$ for the hadronic phase,   
MIT bag model with $B = 85$~MeV/fm$^3$ for the quark phase.   
Results for neutrino-free matter. }
\label{table:latent1} 
\begin{tabular}{lllll}
\hline\noalign{\smallskip}
$T$~~~ & ${\cal Q}$~~~~  & $n_{Q^*}$~~~~  & $n_{H}$~~~~ & $P_0$\\
MeV~~~ & MeV~~~~ & fm$^{-3}$~~~~ & fm$^{-3}$~~~~ & MeV/fm$^3$ \\ 
\noalign{\smallskip}\hline\noalign{\smallskip}
~0~~~     &   ~0.00~~~~ &     0.453~~~~&  0.366~~~~  &     39.95 \\   
~5~~~     &   ~0.56~~~~ &     0.451~~~~&  0.364~~~~  &     39.74 \\  
10~~~     &   ~2.40~~~~ &     0.447~~~~&  0.358~~~~  &     38.58 \\
15~~~     &   ~5.71~~~~ &     0.439~~~~&  0.348~~~~  &     36.55 \\
20~~~     &  10.60~~~~ &      0.428~~~~&  0.334~~~~  &     33.77 \\
25~~~     &  17.17~~~~ &      0.414~~~~&  0.316~~~~  &     30.36 \\
30~~~     &  25.44~~~~ &      0.398~~~~&  0.294~~~~  &    26.53 \\
\noalign{\smallskip}\hline
\end{tabular}
\end{table}
%

\begin{table}
\caption{Same as Table 1, but with trapped neutrinos. }
\label{table:latent3}
\begin{tabular}{l l l l l}
\hline\noalign{\smallskip}
  $T$~~~ & ${\cal Q}$~~~~  & $n_{Q^*}$~~~~  & $n_{H}$~~~~ & $P_0$\\
 MeV~~~ & MeV~~~~ & fm$^{-3}$~~~~ & fm$^{-3}$~~~~ & MeV/fm$^3$ \\ 
\noalign{\smallskip}\hline\noalign{\smallskip}
~0~~~     &   ~0.00~~~~ &     0.603~~~~&  0.516~~~~  &    113.77 \\   
~5~~~     &   ~0.65~~~~ &     0.601~~~~&  0.514~~~~  &    113.11 \\  
10~~~     &   ~2.87~~~~ &     0.594~~~~&  0.509~~~~  &    110.69 \\
15~~~     &   ~6.78~~~~ &     0.580~~~~&  0.499~~~~  &    106.17 \\
20~~~     &  12.65~~~~  &     0.560~~~~&  0.483~~~~  &      99.18 \\
25~~~     &  20.21~~~~  &     0.534~~~~&  0.462~~~~  &      90.12 \\
30~~~     &  29.88~~~~  &     0.502~~~~&  0.434~~~~  &      78.65 \\
\noalign{\smallskip}\hline
\end{tabular}
\end{table}

\section{Quark matter nucleation in cold hadronic stars}
Initially, we assume that the compact star survives the early stages 
of its evolution as a pure hadronic star, and we study quark matter nucleation 
in cold (T = 0) neutrino-free hadronic matter. 
The case of quark matter nucleation at finite temperature in neutrino-free and neutrino-trapped matter 
will be discussed in the Section 6.  

In our scenario, we consider a purely hadronic star whose central pressure  
is increasing due to spin-down or due to mass accretion, {\it e.g.,} from a companion star.     
As the central pressure  exceeds  the deconfinement threshold value $P_0$, 
a virtual drop of quark matter in the Q*-phase can be formed in the central region of the star.   
As soon as a real drop of Q*-matter is formed, it will grow very rapidly 
and the original Hadronic Star will be converted to 
an Hybrid Star or to a Strange Star, depending on the detail of the EOS 
for quark matter employed to model the phase transition. 

In a cold ($T = 0$) and neutrino-free pure hadronic star the formation of the first drop 
of QM  could take place solely via a quantum nucleation process.  
The basic quantity needed to calculate the nucleation time is the energy barrier separating 
the Q*-phase from the metastable hadronic phase.
This energy barrier, which represents the difference in the free energy of the system with and 
without a  Q*-matter droplet, can be written as \cite{lk72,iida98} 
\begin{equation}
  U({\cal R}) = \frac{4}{3}\pi n_{Q^*}(\mu_{Q^*} - \mu_H){\cal R}^3 + 4\pi \sigma {\cal R}^2
\label{eq:potential}
\end{equation}
where ${\cal R}$ is the radius of the droplet (supposed to be spherical), and $\sigma$ is  
the surface tension for the surface separating the hadron from the Q*-phase. 
The energy barrier has a maximum at the critical radius 
${\cal R}_c = 2 \sigma /[n_{Q^*}(\mu_H - \mu_{Q^*})]$.   
We neglected the term associated with the curvature energy and also the terms connected 
with the electrostatic energy, since they are known to only introduce small 
corrections \cite{iida98,bo04}.  
The value of the surface tension $\sigma$ for the interface separating the quark and hadron phase 
is poorly known, and typically values used in the literature range within $10-50$~MeV fm$^{-2}$ 
\cite{hei93}). 
Larger values of $\sigma$ in the range 50--160~MeV fm$^{-2}$ have also been obtained in the 
literature \cite{vosk+03,LGA13}. Clearly these large values of the surface tension disfavor or inhibit quark matter nucleation in hadronic stars \cite{lug+11}. 
  
\begin{figure}
\vspace{0.5cm}
\includegraphics[scale=0.30, angle = 270 ]{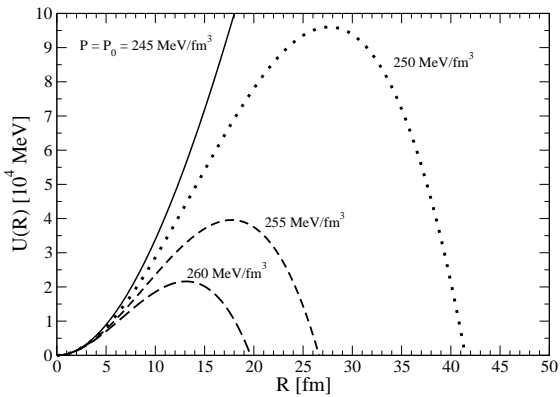}
\caption{Potential energy of the QM drop as a function of the radius of the drop for several pressures above $P_0$ at $T = 0$.  
The hadronic phase is described with the GM3 model whereas for the Q$^*$ phases is employed the MIT bag 
model with $m_s=150$ MeV, $B=152.45$ MeV/fm$^3$. 
The surface tension $\sigma$ is taken equal to $30$ MeV/fm$^2$. } 
\label{barrier_T0}
\end{figure}

The nucleation time, {\it i.e.} the time needed to form the first drop of the Q*-phase, 
can be straightforwardly evaluated within a semi-classical approach \cite{iida98,be02,be03}. 
First one computes, in the Wentzel -Kramers -Brillouin (WKB) approximation, the ground state 
energy $E_0$ and the oscillation frequency $\nu_0$ of the drop in the potential well $U({\cal R})$. 
Then, the probability of tunneling is given by 
\begin{equation}
  p_0=exp\left[-\frac{A(E_0)}{\hbar}\right]
\label{eq:prob}
\end{equation}
where $A$ is the action under the potential barrier which in a relativistic framework reads
\begin{equation}
A(E) = \frac{2}{c} 
\int_{{\cal R}_-}^{{\cal R}_+}\sqrt{[2{\cal M}({\cal R})c^2 +E-U({\cal R})][U({\cal R})-E]}d{\cal R} \ ,
\label{eq:action}
\end{equation}
with ${\cal R}_\pm$ the classical turning points and
\begin{equation}
 {\cal M}({\cal R}) = 4\pi \rho_H\left(1-\frac{n_{Q^*}}{n_H}\right)^2 {\cal R}^3
\label{eq:mass}
\end{equation}
the droplet effective mass, with $\rho_H$ and $n_H$ the hadron energy density and the hadron baryon 
number density, respectively. The nucleation time is then equal to
\begin{equation}
  \tau_q =  (\nu_0 p_0 N_c)^{-1} \ ,
\label{eq:time}
\end{equation}
where $N_c \sim 10^{48}$  is the number of nucleation centers expected in the innermost part 
($r \leq R_{nuc} \sim 100$ m) of the HS, where the pressure and temperature 
(when we will consider the finite T case) can be considered constant and equal to their 
central values.    
The uncertainty in the value of $N_c$ is expected to be within one or two orders of magnitude. 
In any case, all the qualitative features of our scenario will not be affected by this 
uncertainty \cite{be02,be03,bo04}.  

As a consequence of the surface effects it is necessary to have an overpressure $\Delta P= P-P_0 > 0$ with respect to the bulk transition point $P_0$ to create a drop of deconfinement quark matter in the hadronic environment. The higher the overpressure, the easier to nucleate the first drop of $Q^*$ 
matter. In other words, the higher the mass of the metastable pure hadronic star, the shorter the time to nucleate a quark matter drop at the center of the star.   

As an illustrative example, we plot in Fig.\ \ref{barrier_T0} the potential energy $U(\R)$ 
for the formation of a quark matter droplet for different values of the stellar 
central pressure $P_c$ above the  deconfinement threshold value $P_0$.   
The curves in Fig.\ \ref{barrier_T0} are relative to  a given set of EOS for the two phases of 
dense matter and  to a fixed value of  the surface tension $\sigma$ (see figure caption).    
As expected the potential barrier is lowered as central pressure increases. 

\begin{figure}
\vspace{0.5cm}
\resizebox{0.43\textwidth}{!}{
\includegraphics{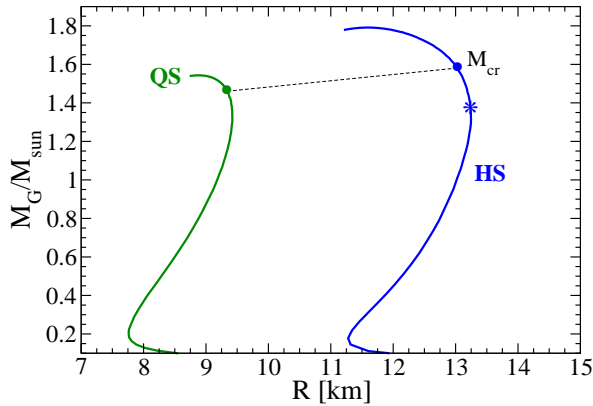} }
\caption{(Color on line) Mass-radius relation for pure hadronic star (HS) and hybrid star (QS) 
configurations.  
The configuration marked with an asterisk represents the HS for which 
$\tau_q = \infty$ ({\it i.e.} $P_c = P_0$). The conversion process of the HS, with a gravitational 
mass equal  to $M_{cr}$,  into the final QS is denoted by the full circles connected by a dashed line.
Results are relative to the GM1 model with $x_{\sigma} = 0.6$  for the hadronic phase 
and the MIT bag model EOS with B = 85~MeV/fm$^3$ for the quark phase.   
The surface tension is   $\sigma =  30~{\rm MeV/fm}^2$. 
Stellar masses are in unit of the mass of the sun, $M_{sun} = 1.989 \times 10^{33}$~g. } 
\label{MR}
\end{figure}

Thus a pure hadronic star, having  a central pressure $P_c$ larger than the transition 
pressure $P_0$ for the formation of the Q*-phase, is metastable \cite{be02,be03,bo04} 
to the ``decay'' (conversion) to a quark star (QS) {\it i.e.} to a stellar configuration 
in which deconfined quark matter is present. 

These metastable HSs  have a {\it mean-life time}  which is related to the nucleation time to 
form the first critical-size drop of deconfined matter in their interior 
(the actual  {\it mean-life time} of the HS will depend on the 
mass accretion or on the spin-down rate which modifies the nucleation time via an explicit 
time dependence of the stellar central pressure).  Following Refs.~\cite{be02,be03,bo04} 
we define as {\it critical  mass} $M_{cr}$ of the metastable HSs,    
the value of the  gravitational mass for which the nucleation time is equal to one year: 
$M_{cr} \equiv M^{HS}(\tau_q = 1 {\rm yr})$.   
Pure hadronic stars with $M^{HS} > M_{cr}$ are very unlikely to be observed.   
Thus  $M_{cr}$  plays the role of an {\it effective maximum mass} \cite{bo04} 
for the hadronic branch of compact stars. Notice that the Oppenheimer--Volkoff \cite{ov39} 
maximum mass $M^{HS}_{max}$ is determined by the overall stiffness of the EOS for 
hadronic matter,  whereas the value of $M_{cr}$  will depend in addition on the bulk 
properties of the EOS for quark matter and on the properties at the interface between 
the confined and deconfined phases of matter ({\it e.g.,} the surface tension $\sigma$).

\begin{figure*}[t]
\begin{center}
\resizebox{0.58\textwidth}{!}
{%
\includegraphics[clip=true]{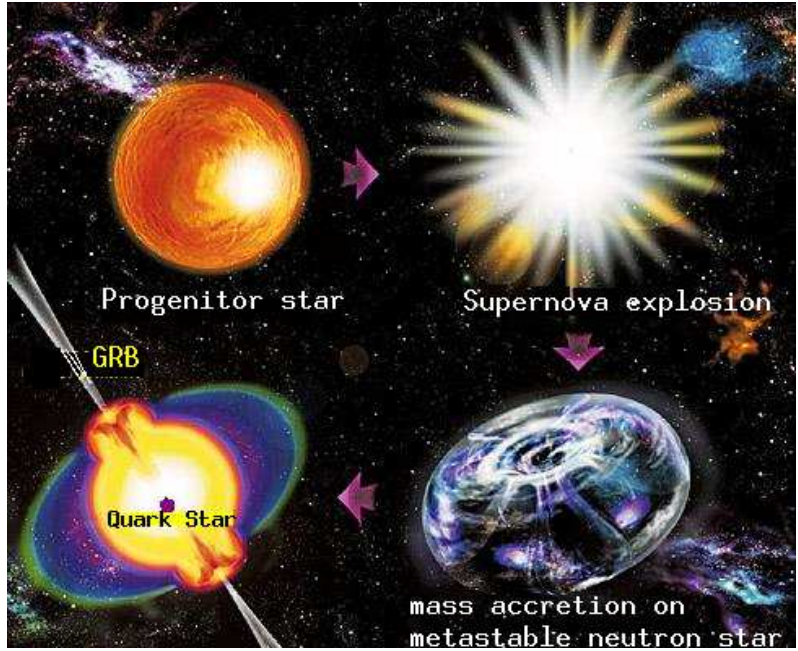} 
}
\vspace{0.2cm}
\caption{(Color on line) A schematic representation of the last stages of the evolution of a 
massive star ($M > 8~M_{sun}$) leading to the delayed conversion of a pure hadronic star 
to a quark star (hybrid or strange star) and to the emission of of a neutrino burst 
and possibly to a gamma ray burst \cite{be02,be03}.  
Clockwise from the upper left corner of the figure: 
(i) nuclear burning stage of the progenitor star; 
(ii) supernova explosion and birth of a pure hadronic star ("neutron star"); 
(iii) mass accretion on the metastable hadronic star; 
(iv) conversion process of the hadronic star to a quark star (second "explosion") neutrino burst and 
gamma ray burst.}
\label{QDN} 
\end{center}
\end{figure*}

These findings are exemplified in  Fig.\ \ref{MR}, where we show  the mass-radius (MR) 
curve for hadronic stars (HS) and that for quark stars (QS). 
The configuration marked with an asterisk on the hadronic MR curve represents 
the HS for which the central pressure is equal to $P_0$ and thus $\tau_q = \infty$.     
The full circle on the HS sequence represents the critical mass configuration $M_{cr}$,  
in the case  $\sigma = 30$ MeV/fm$^2$.  
The  full circle on the QS mass-radius curve represents the hybrid star  
which is formed from the conversion of the hadronic star with $M^{HS} = M_{cr}$. 
We assume \cite{grb} that during the stellar conversion process 
the total number of baryons in the star (or in other words the stellar baryonic mass $M_B$)   
is conserved. Thus the total energy liberated in the stellar conversion is given by \cite{grb} 
the difference between the gravitational mass of the initial hadronic star    
($M_{in} \equiv M_{cr}$) and that of the final quark star $M_{fin}$ configuration with 
the same baryonic mass ({\it i.e.} with $M_{B,cr} = M_{B,fin}\,$):    
\begin{equation}
          E_{conv} = (M_{in} - M_{fin}) c^2  \,. 
\label{Econv}
\end{equation}

It has been shown \cite{grb,be02,be03,bo04,lug05,blv07,bppv08} 
(see also Tabs. \ref{GM1_B_a4_sig_10} and \ref{GM1_B_a4_sig_30})  
that $E_{conv} = 0.5\,$--$\,4.0 \times 10^{53}~\mathrm{erg}$. 
This huge amount of released energy will cause a powerful neutrino burst, likely accompanied by intense gravitational waves emission, and conceivably it could cause a second delayed explosion{\footnote{delayed with respect to the first explosion, i.e. the supernova explosion, which formed the hadronic star ("neutron star"). In other words, we assume (in this section) that quark matter is not formed during the stellar collapse generating the supernova explosion and the protohadronic star. This possibility well be discussed 
in Section 6.}.} Under favorable physical conditions this second explosion could be the energy source of a powerful gamma-ray burst (GRB) \cite{be02,be03}. Thus this scenario is able to explain a "delayed" connection between supernova explosions and GRBs.  

It has also been suggested \cite{bom-popov04} that the delayed stellar conversion 
process of a pure HS to a QS, can impart a second kick to the nascent QS with respect to the first kick imparted to the newly  formed HS during the supernova explosion. 
Thus this model \cite{bom-popov04} could explain in a natural way the observed bimodal distribution 
of the kick velocities of radio pulsars \cite{arzou02}. 
Thus, according to the authors of Ref.~\cite{bom-popov04}, the low velocity component of the 
pulsar velocity distribution receives contributions mainly from hadronic stars which have 
passed through a single explosion (the supernova explosion). 
The high velocity component is mostly composed of quark stars which have received a second kick 
due to the energy release associated to the stellar conversion process.  

The last stages of the evolution of a massive star ($M > 8~M_{sun}$), within the scenario proposed 
by the authors of Ref.~\cite{be02,be03}, are schematically depicted in Fig.\ \ref{QDN}. 

\begin{figure*}[t]
\begin{center}
\resizebox{0.70\textwidth}{!}
{%
\includegraphics[clip=true]{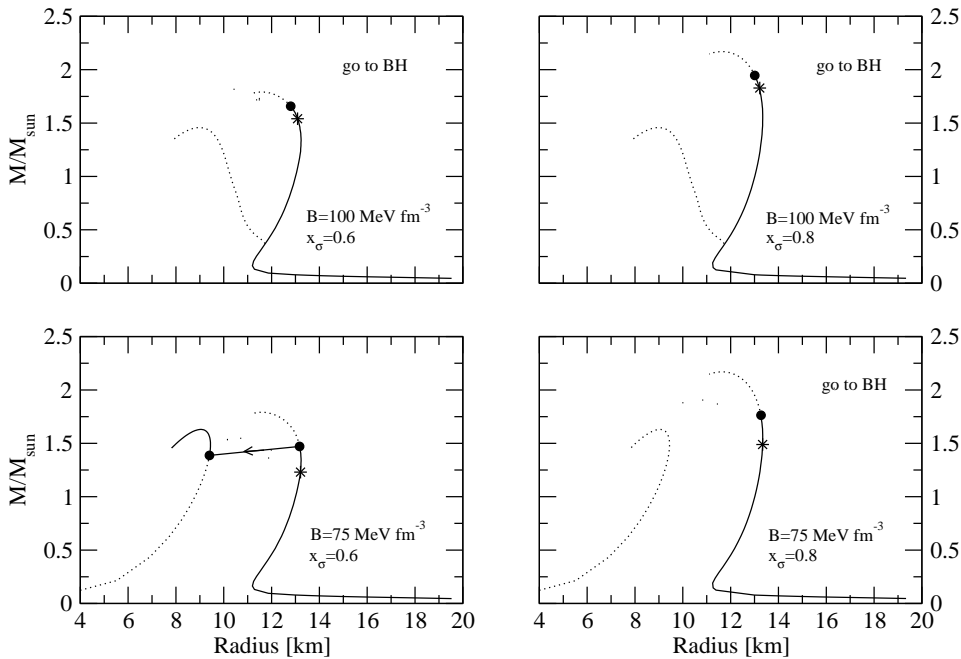}
}
\caption{Mass-radius relation for a pure HS described within the GM1 parametrization 
 and that of the hybrid stars or strange stars configurations for two values of the bag constant
 ($B=75$ and 100 MeV/fm$^3$) and two values of the hyperon-meson coupling 
 ($x_\sigma=0.6,$ and 0.8) and $m_s=150$ MeV. 
 The configuration marked with an asterisk represents 
 in all cases the HS for which the central pressure is equal to $P_0$. 
 The conversion process of the HS, with a gravitational mass equal to $M_{cr}$, into
 a final hybrid star or strange star is denoted by the full circles connected by an arrow. 
 The label "go to BH" designates the case in which the critical mass HS collapses to a black hole. 
 In all the panels $\sigma$ is taken equal to 30 MeV/fm$^2$.}
\label{GM1_MR}      
\end{center}
\end{figure*}

In Fig.\ \ref{GM1_MR} we show the mass-radius relation in the case of the GM1 EOS for two 
different values of the hyperon coupling ($x_\sigma = 0.6$ and 0.8) 
and for two different values of the bag constant ($B = 75$ and 100~MeV/fm$^3$).  
These results illustrate that the outcome of the scenario proposed in Ref.~\cite{be02,be03},  
and, in particular, the final fate of the critical mass HS, depends on the details of the EOS describing 
the two matter phases. 
Specifically, as shown in Fig.\ \ref{GM1_MR}, for some values of the EOS parameters, 
the critical mass HS will collapse to a black hole (BH) for the reason that the baryonic mass of the 
critical mass configuration is larger than the maximum baryonic mass for the quark star sequence 
({\it i.e.} $M_{B,cr} > M_{B,max}^{QS} $).

In Tables 3 and 4, we report the calculated values of the critical gravitational (baryonic) mass 
$M_{cr}$ ($M_{B,cr}$), the value of mass of the final QS configuration $M_{fin}$ 
and the energy $E_{conv}$ released in the stellar conversion process.    
The results are relative to the GM1 model with $x_{\sigma} = 0.7$ for the hadronic phase 
and to the extended bag model EOS of Ref.~\cite{Fra01,Alf05,weis11} for the quark phase,  
using different values for the effective bag constant $B_{eff}$ and the perturbative  
QCD correction term $a_4$. 
The results in Tab. 3 are relative to a surface tension $\sigma = 10~{\rm MeV/fm}^2$, 
whereas those in Tab. 4 to $\sigma = 30~{\rm MeV/fm}^2$. 
Notice that for these EOS models and for the parameters reported in Tables 3 and 4, both 
the critical mass of the hadronic star sequence and the maximum mass $M_{max}^{QS}$ 
of the quark star sequence are  consistent with present measured neutron star masses and, in particular, 
with the mass $M = 1.97 \pm 0.04 \, M_{sun}$ of PSR~J1614-2230 \cite{demo10}  
and $M = 2.01 \pm 0.04 \, M_{sun}$ of PSR~ J0348+0432 \cite{anto13}. 

\begin{table}
\caption{Critical mass and energy released in the conversion process of an HS into a QS.  
Results are relative to the GM1 model with $x_{\sigma} = 0.7$ for the hadronic phase 
and the extended bag model EOS of Ref.~\cite{Fra01,Alf05,weis11} for the quark phase,  
using different values for the effective bag constant $B_{eff}$ and the perturbative 
QCD correction term $a_4$. For the quark masses we use $m_s = 100~\rm{MeV}$, $m_u = m_d=0$. 
The value of the critical gravitational (baryonic) mass of the HS sequence is reported on the 
column labeled $M_{cr}$ ($M_{B,cr}$), whereas those of the mass of the final QS formed in the stellar conversion process of the critical mass HS are shown on the column labeled $M_{fin}$. 
The column labeled $M_{max}^{QS}$ denotes the maximum gravitational mass of the QS sequence.  
Finally the energy released in the stellar conversion process are shown on the column 
labeled $E_{conv}$. 
Units of $B_{eff}$ and $\sigma$ are MeV/fm$^3$ and MeV/fm$^2$  respectively.  
All stellar masses are given in units of the mass of the sun, $M_{sun} = 1.989 \times 10^{33}$~g, 
and $E_{conv}$ is given in units of $10^{53}$ erg. 
The surface tension is $\sigma = 10~{\rm MeV/fm}^2$.}
\label{GM1_B_a4_sig_10}
\begin{tabular}{l l l l l l l}
\hline\noalign{\smallskip}
  $B_{eff}$   & $a_4$ & $M_{cr}$    & $M_{B,cr}$   & $M_{fin}$ &$M_{max}^{QS}$  & $E_{conv}$  \\ 
\noalign{\smallskip}\hline\noalign{\smallskip}

  37.63 & 0.65&  2.003 & 2.312 & 1.848   & 2.318  & 2.77 \\
  50.72 & 0.65&  2.033 & 2.354 & 1.954   & 2.013  & 1.41 \\
  41.14 & 0.70&  1.891 & 2.158 & 1.731   & 2.229  & 2.86 \\
  47.20 & 0.70&  1.940 & 2.225 & 1.814   & 2.088  & 2.25 \\
\noalign{\smallskip}\hline
\end{tabular}
\end{table}

\begin{table}
\caption{Same as Table 3, but for a value of the surface tension $\sigma = 30$~MeV/fm$^2$. }
\label{GM1_B_a4_sig_30}
\begin{tabular}{l l l l l l l}
\hline\noalign{\smallskip}
 $B_{eff}$   & $a_4$ & $M_{cr}$    & $M_{B,cr}$   & $M_{fin}$ &$M_{max}^{QS}$  & $E_{conv}$ \\ 
\noalign{\smallskip}\hline\noalign{\smallskip}
  
 37.63  & 0.65&  2.021 & 2.328 & 1.859   & 2.318  &  2.90 \\
 50.72  & 0.65&  2.039 & 2.362 & 1.960   & 2.013  &  1.41 \\
 41.14  & 0.70&  1.995 & 2.301 & 1.832   & 2.229  &  2.91 \\
 47.20  & 0.70&  1.973 & 2.270 & 1.846   & 2.088  &  2.27 \\ 
\noalign{\smallskip}\hline
\end{tabular}
\end{table}

\begin{figure*}[t]
\begin{center}
\resizebox{0.50\textwidth}{!}
{%
\includegraphics[clip=true]{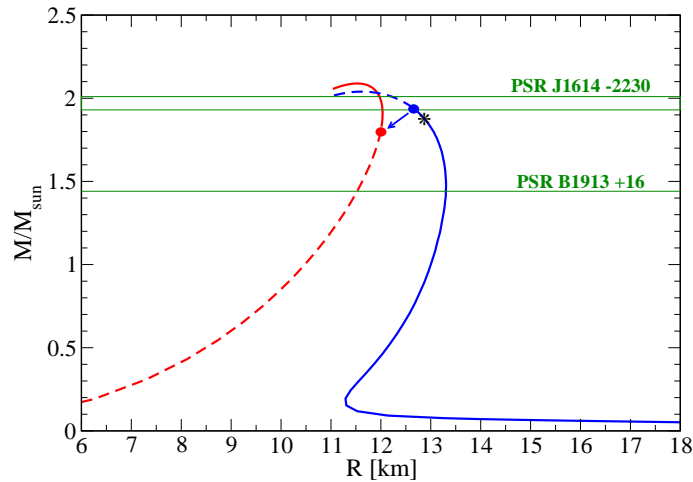}
}
\caption{(Color on line) Mass-radius relation for a pure HS described within the GM1 model 
of hyperonic matter with hyperon-$\sigma$ meson coupling $x_\sigma = 0.7,$  
and for strange star configurations with the extended bag model EOS of Ref.~\cite{Fra01,Alf05,weis11} 
with $B_{eff} = 47.2~\mathrm{MeV/fm}^3$ and $a_4 = 0.7$. 
The configuration marked with an asterisk represents the HS for which 
$\tau_q = \infty$.   
The conversion process of the HS, with a gravitational mass equal to $M_{cr}$, 
into the final QS is denoted by the full circles connected by an arrow. 
The values of the critical gravitational (baryonic) mass and of the final QS mass  
are calculated for a surface tension $\sigma = 10~{\rm MeV/fm}^2$. 
Their numerical values are given in the last row of Tab. 3.
The lower horizontal line represents the mass $M = 1.4398 \pm 0.0002 \, M_{sun}$ \cite{HT75} 
of the pulsar PSR~B1913+16, whereas the higher horizontal lines represent 
the mass $M = 1.97 \pm 0.04 \, M_{sun}$ of PSR~J1614-2230 \cite{demo10}.} 
\label{MR_GM1_xs07_B_138_a4_08} 
\end{center}
\end{figure*}

The mass-radius relation for the stellar configurations relative to the entry of the 
last row of Tab. \ref{GM1_B_a4_sig_10} is presented in Fig.~\ref{MR_GM1_xs07_B_138_a4_08}.  
As we can see, for this EOS parametrization, PSR~B1913+16 
(which has a mass $M = 1.4398 \pm 0.0002 \, M_{sun}$ \cite{HT75}) can be interpreted as 
a pure HS,  whereas PSR~J1614-2230 is more likely a QS.    

The stellar conversion process, described so far, will start to populate 
the new branch of quark stars, {\it i.e.} the part of the QS sequence above the full circle 
(see Fig.\ \ref{MR} and \ref{MR_GM1_xs07_B_138_a4_08}).  
Long term accretion on the QS can next produce stars with masses up to the maximum 
mass $M^{QS}_{max}$ for the quark star configurations. 
Thus within this scenario one has two coexisting families of compact stars: pure hadronic stars 
and quark stars \cite{bo04} (see \cite{drago15} in this Topical Issue). 
The quark star branch is occasionally referred to as the ``third family'' of compact stars, 
considering white dwarfs as the first family and pure hadronic stars as the second family. 
Notice also that there is a range of values of stellar gravitational mass 
(see Fig.\ \ref{MR} and \ref{MR_GM1_xs07_B_138_a4_08}) where hadronic stars 
and quark stars with the same gravitational mass can exist (``twin stars'').   

\section{The limiting mass of compact stars: extending the Oppenheimer-Volkoff mass limit concept} 
The possibility to have metastable hadronic stars, together with the predicted coexistence 
of two distinct families of compact stars, demands an extension of the 
concept of maximum mass of a ``neutron star'' with respect to 
the {\it classical} one introduced by Oppenheimer and Volkoff in 1939 \cite{ov39}.      
Since metastable HS with a ``short'' {\it mean-life time} are very unlikely to be observed,  
the extended concept of maximum mass must be introduced in view of the comparison 
with  the values of the mass of compact stars deduced from direct astrophysical 
observation.  
Having in mind this operational definition, the authors of Ref. \cite{bo04} 
called {\it limiting mass} of a compact star, and denoted it as $M_{lim}$, 
the physical quantity defined in the following way: 

\noindent 
({\it a})  if the nucleation time $\tau(M^{HS}_{max})$  associated to the maximum mass 
configuration for the hadronic star sequence is of the same order or much larger  than  
the age of the universe $T_{univ}$,   then 
\be
         M_{lim}  =  M^{HS}_{max} \, ,  
\ee
in other words, the limiting mass in this case coincides with the Oppenheimer - Volkoff 
maximum mass for the hadronic star sequence. 

\noindent 
({\it b}) If the critical mass $M_{cr}$  is smaller than $M^{HS}_{max}$  
({\it i.e.}  $\tau(M^{HS}_{max}) < 1$~yr), 
thus the limiting mass for compact stars is equal to the largest value between the 
critical mass for the HS and the maximum mass for the quark star (HyS or SS) sequence 
\be
         M_{lim} =   max \big[M_{cr} \, ,  M^{QS}_{max} \big] \, .
\ee

\noindent 
({\it c}) Finally, one must consider an ``intermediate'' situation for which      
                          $1 {\rm yr} <  \tau(M^{HS}_{max})  <  T_{univ}$.  
As the reader can easily  realize, now 
\be
         M_{lim} =   max \big[M^{HS}_{max} \, ,  M^{QS}_{max} \big] \, , 
\ee
depending on the details of the EOS which could give $M^{HS}_{max} > M^{QS}_{max}$  
or vice versa.

\section{Quark matter nucleation in proto-hadronic stars} 

A neutron star at birth (proto-neutron star) is very hot (T =  10 -- 30 MeV) 
with neutrinos being still trapped in the stellar  
interior \cite{BurLat86,bom95,prak97,pons99,vid+03,marg+03}.      
Subsequent neutrino diffusion causes deleptonization and heats the stellar matter to 
an approximately uniform entropy per baryon $\tilde {S}$~=~1 -- 2 (in units of the Boltzmann's 
constant $k_B$).  Depending on the stellar composition, during this stage 
neutrino escape can lead the more ``massive'' stellar configurations to the formation 
of a black hole \cite{bomb96,prak97}.  However, if the mass of the star is sufficiently small, 
the star will remain stable and it will cool to temperatures well below 1~MeV  within 
a cooling time $t_{cool} \sim$~a~few~$10^2$~s, as the neutrinos continue to carry energy away 
from the stellar material \cite{BurLat86,prak97,pons99}.
Thus in a proto-neutron star, the quark deconfinement phase transition will be likely triggered 
by a  thermal nucleation process \cite{ho92,ho94,ol94,harko04}. In fact, for sufficiently high  temperatures, thermal nucleation is a much more efficient process with respect to the 
quantum nucleation mechanism. 

In Ref. \cite{bom+09,bom+11} we established the physical conditions under which a newborn hadronic star 
(proto-hadronic star, PHS) could survive the early stages of its evolution without "decaying" 
to a quark star.  

\begin{figure}
\vspace{0.5cm}     
\resizebox{0.41\textwidth}{!}{%
 \includegraphics{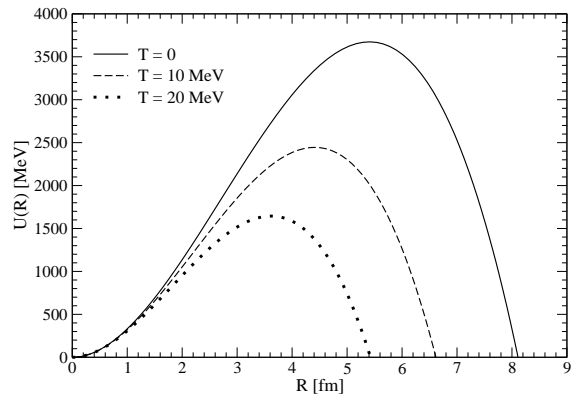} }
\caption{Energy barrier for a virtual drop of the Q*-phase  in $\beta$-stable neutrino-free hadronic 
matter as a function of the droplet radius and for different temperatures for a fixed pressure 
$P = 57$~MeV/fm$^3$.  
Results are relative to the GM1$_{0.6}$--B85 equation of state  and  
surface tension $\sigma =  30~{\rm MeV/fm}^2$.} 
\label{barrier_T} 
\end{figure}

\begin{figure}
\vspace{0.5cm} 
\resizebox{0.42\textwidth}{!}{%
 \includegraphics{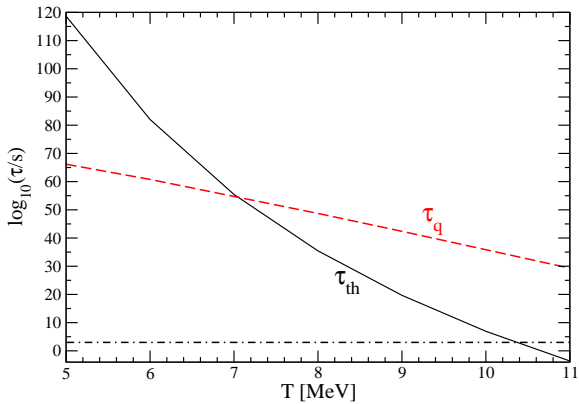} }
\caption{(Color on line) Thermal ($\tau_{th}$) and quantum ($\tau_q$) nucleation time of quark matter (Q*-phase) 
in $\beta$-stable neutrino-free  hadronic matter as a function of temperature at fixed 
pressure $P= 57$~MeV/fm$^3$.  The crossover temperature is  $T_{co}=7.05$~MeV. 
The limiting conversion temperature for the proto-hadronic star is, in this case, $\Theta = 10.3$~MeV, obtained from the intersection of the thermal nucleation time curve (continuous line) 
and the dot-dashed line representing  $\log_{10}(\tau/{\rm s}) = 3$. 
The surface tension is   $\sigma =  30~{\rm MeV/fm}^2$. 
Results are relative to GM1$_{0.6}$--B85 EOS.} 
\label{nucl_time} 
\end{figure}  

According to the Langer theory \cite{lang69,LanTur73} of homogeneous nucleation 
the thermal nucleation rate can be written \cite{LanTur73} as  
\begin{equation}
     I =\frac{\kappa}{2 \pi} \Omega_0 \exp (- U({\cal R}_c, T) /T)
\label{eq:therm_rate}
\end{equation}
where $\kappa$ is the so-called dynamical prefactor, which is related to the growth rate of 
the drop radius $\cal R$ near the critical radius (${\cal R}_c$), and  $\Omega_0$ is 
the so-called statistical prefactor, which  measures the phase-space volume of the 
saddle-point region around ${\cal R}_c$. 
We have used \cite{bom+09,bom+11} for $\kappa$ and  $\Omega_0$ the expressions derived in 
Refs.~\cite{CseKap92,VenVis94}, where the Langer nucleation theory has been extended 
to the case of first order phase transitions occurring in relativistic systems, as in the case of the quark deconfinement transition.  
The dominant factor in the nucleation rate (\ref{eq:therm_rate}) is the exponential, in which 
$U({\cal R}_c, T)$ is the activation energy, {\it i.e.} the change in the free energy of the system 
required to activate the formation of a critical size droplet. 
This quantity has been calculated \cite{bom+09,bom+11} using the generalization of 
Eq.~(\ref{eq:potential})  to the case of $T \neq 0$ (thus using finite temperature EOS for the hadronic and the quark phases), and assuming a temperature independent surface tension.   

The thermal nucleation time $\tau_{th}$, relative to the innermost stellar region 
($V_{nuc} = (4 \pi/3) R_{nuc}^3$, with $R_{nuc} \sim 100$ m) where almost constant pressure 
and temperature occur, can thus be written as   
\begin{equation}
  \tau_{th} = (V_{nuc} \, I )^{-1} \ . 
\label{eq:tau_th}
\end{equation} 

\begin{figure}
\vspace{0.5cm}
\resizebox{0.42\textwidth}{!}{%
 \includegraphics{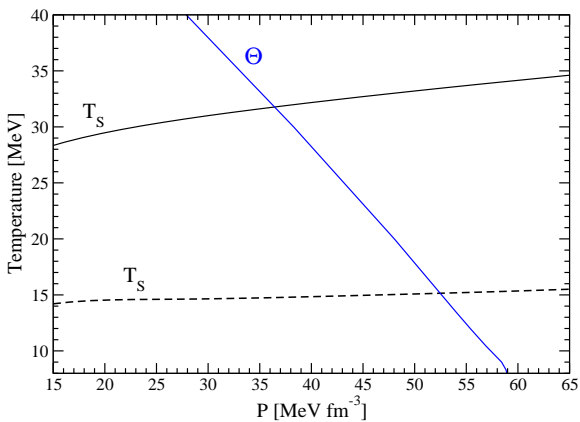} }
\caption{(Color on line) 
The limiting conversion temperature $\Theta$ for a newborn hadronic star as a function 
of the  central stellar pressure. 
Newborn hadronic stars with a central temperature and pressure located on the right side of 
the curve $\Theta(P)$ will nucleate a Q*-matter drop during the early stages of their evolution,  
and will finally evolve to cold and deleptonized quark stars, or will collapse to  black holes.  
The lines labeled  $T_S$ represent the stellar matter temperature as a function of pressure 
at fixed entropies per baryon $\tilde S/k_B = 1$ (dashed line) and $2$ (solid line). 
Results are relative to GM1$_{0.6}$--B85 EOS.}
\label{theta_B85} 
\end{figure}  

\begin{figure}
\vspace{0.5cm}
\vspace{0.6cm}
\resizebox{0.42\textwidth}{!}{%
 \includegraphics{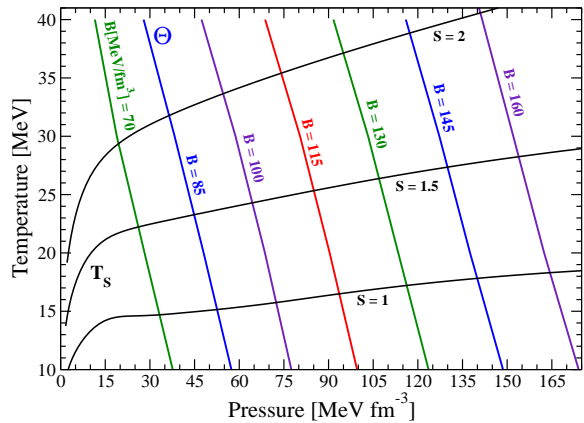} }
\caption{(Color on line) The limiting conversion temperature $\Theta$ for a newborn hadronic star 
as a function of the  central stellar pressure for different values of the bag constant $B$.  
The lines labeled  $T_S$ represent the stellar matter temperature 
as a function of pressure at fixed entropies per baryon $\tilde S/k_B = 1$, $1.5$, $2$.  
Results for neutrino-free matter.}   
\label{theta_GM1}  
\end{figure}  

In Fig.\ \ref{barrier_T}, we represent the energy barrier for a virtual drop of the Q*-phase 
in the neutrino-free hadronic phase as a function of the droplet radius and for different temperatures at a fixed  pressure $P = 57$~MeV/fm$^3$.  Results in Fig.\ \ref{barrier_T} 
are obtained using the GM1$_{0.6}$--B85 equation of state.  
As expected, from the results plotted in Fig.\ \ref{gibbs_GM1_B85},  the energy 
barrier $U({\cal R}, T)$ and the droplet critical radius  ${\cal R}_c$ decrease as the matter 
temperature is increased. This effect favors the Q*-phase formation and, in particular,  
increases (decreases) the quantum nucleation rate (nucleation time $\tau_q$) with respect 
to the corresponding quantities  calculated at  $T=0$.   

In Fig.\ \ref{nucl_time}, we plot the quantum and thermal nucleation times of the  
Q*-phase  in $\beta$-stable neutrino-free hadronic matter as a function of temperature 
and at a fixed pressure $P=57$~MeV/fm$^3$.   
As expected, we find a crossover temperature $T_{co}$ above which thermal nucleation is 
dominant with respect to the quantum nucleation mechanism.  
For the case reported in Fig.\ \ref{nucl_time}, we have $T_{co}=7.05$~MeV and the 
corresponding nucleation time is $\log_{10}(\tau/{\rm s}) = 54.4$. 

Having in mind the physical conditions in the interior of a PHS \cite{BurLat86,prak97}, 
to establish if this star will survive the early stages of its evolution 
without decaying to a quark star, one has to compare the quark matter nucleation time 
$\tau=\min(\tau_q,\tau_{th})$ with the cooling time  $t_{cool} \sim$~a~few~$10^2$~s.   
If $\tau >> t_{cool}$ then quark matter nucleation will not likely occur 
in the newly formed star, and this star will evolve to a cold deleptonized configuration.  
We thus introduce the concept of {\it limiting conversion temperature} $\Theta$ for the  
proto-hadronic star and define it as the value of the stellar central temperature $T_c$ for which 
the Q*-matter nucleation time is  equal to $10^3$~s. The limiting conversion temperature 
$\Theta$ will clearly depend on the value of the stellar central pressure (and thus on the value 
of the stellar mass).  

The limiting conversion temperature $\Theta$ is plotted in Fig.\ \ref{theta_B85} as a 
function of the stellar central pressure. A proto-hadronic star with a central temperature  
$T_c > \Theta$  will likely nucleate a Q*-matter drop during the early stages of its evolution,  
and will finally evolve to a cold and deleptonized quark star, or will collapse to a black hole 
(depending on the value of the stellar baryonic mass $M_B$ and on the EOS).    

In  Fig.\ \ref{theta_GM1} we plot the limiting conversion temperature $\Theta$ 
for a newborn HS for different values of the bag constant. The increase in B 
produces a growth of the region of the $P$--$T$ plane where the proto-hadronic star could 
survive Q* nucleation and thus evolve to a cold hadronic star.  

For an isoentropic stellar core \cite{BurLat86,prak97}, the central temperature of the 
proto-hadronic star is given, for the GM1$_{0.6}$ EOS model, by the lines labeled by $T_S$ 
in Fig.\ \ref{theta_B85},  relative to the case $\tilde {S} = 1~k_B$ (dashed curve) and $\tilde {S} = 2~k_B$ (continuous curve).  
The intersection point ($P_S,\Theta_S$) between the two curves $\Theta(P)$ and $T_S(P)$
thus gives the central pressure and temperature of the configuration that we denote as 
the {\it critical mass} configuration of the proto-hadronic stellar sequence.
Taking $\sigma =  30~{\rm MeV/fm}^2$ and $\tilde {S} = 2~k_B$ we get  
$M_{cr} = 1.390~M_{sun}$ for the gravitational critical mass  and $M_{B,cr} = 1.492~M_{sun}$ 
for the  baryonic critical mass (being $M_{sun} = 1.989 \times 10^{33}$~g the mass of the sun).   

\begin{figure}
\vspace{0.5cm}
\resizebox{0.43\textwidth}{!}{%
\includegraphics{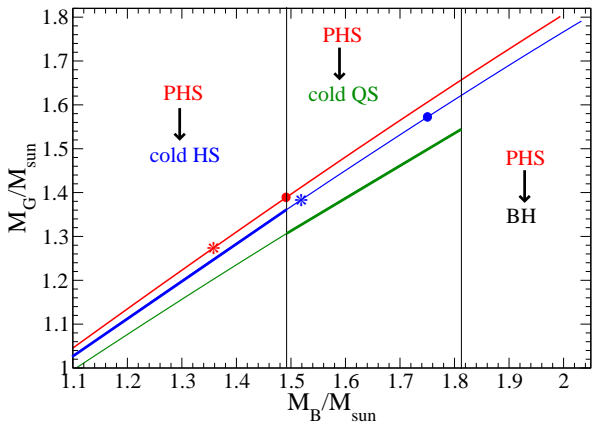} }
\vspace{0.2cm}
\caption{(Color on line) Stellar equilibrium sequences in the gravitational--baryonic mass plane for the 
GM1$_{0.6}$--B85 EOS and $\sigma =30$~MeV/fm$^2$.  
The upper (red) line represents the PHS sequence ($\tilde {S} = 2~k_B$). 
The middle (blue) line represents the cold HS sequence. 
The asterisk and the full circle on these lines represent respectively the stellar configuration with 
$\tau = \infty$ and the critical mass configuration $M_{cr}$. 
The lower (green) line represent the cold QS sequence.  
Assuming  $M_B =~$const, the evolution of a PHS in this plane occurs along a vertical line.} 
\label{Mg_Mb_GM1_B85_S2} 
\end{figure}  

The evolution of a PHS within our scenario is delineated in Fig.\ \ref{Mg_Mb_GM1_B85_S2},  
where we plot the appropriate stellar equilibrium sequences in the gravitational--baryonic mass plane for the GM1$_{0.6}$--B85 EOS  and $\sigma =  30~{\rm MeV/fm}^2$.  
The upper (red) line represents the PHS sequence, {\it i.e.} isoentropic HSs ($\tilde {S} = 2~k_B$) 
and neutrino-free matter. The middle (blue) line represents the cold HS sequence. 
The asterisk and the full circle on these lines represent respectively the stellar configuration 
with $\tau = \infty$ and the critical mass configuration. Finally, the lower (green) line 
represent the cold QS sequence. 
We assume $M_B =$~const during these stages of the stellar evolution.   
Thus according to the results in Fig.\ \ref{Mg_Mb_GM1_B85_S2}, proto-hadronic stars 
with a baryonic mass $M_B < 1.492~M_{sun}$ will survive Q*-matter {\it early nucleation}  
({\it i.e.} nucleation within the cooling time $t_{cool} \sim$~a~few~$10^2$~s) and in the end 
they will form stable ($\tau = \infty$) cold hadronic stars.    
Proto-hadronic stars with $1.492~M_{sun} \le  M_B < 1.813~M_{sun}$ (the maximum baryonic mass 
$M_{B,max}^{QS}$ of the cold QS sequence for the present EOS) will experience early nucleation 
of a Q*-matter drop and will ultimately form a cold deleptonized quark star.  
The last possibility is for PHSs having  $M_B > 1.813~M_{sun}$. In this case the early nucleation 
of a Q*-matter drop will trigger a stellar conversion process  to a cold QS configuration with 
$M_B > M_{B,max}^{QS} \,$, thus these PHSs will finally form black holes. 

The outcomes of this scenario are not altered by neutrino trapping effects in 
hot $\beta$-stable hadronic matter. 
In fact, for proto-hadronic stars with trapped neutrinos ($\nu$PHSs) (with a lepton fraction 
$Y_L = 0.4$ and $\tilde {S} = 2~k_B$) we find \cite{bom+11} (for the same EOS and 
surface tension $\sigma$ used in Fig.\ \ref{Mg_Mb_GM1_B85_S2}) a critical baryonic 
mass  $M_{B,cr} = 1.96~M_{sun}$. 
Thus $\nu$PHSs with $M_B \geq M_{B,cr}$ after neutrino escape and cooling will finally evolve to black holes. The fate of a $\nu$PHS with  $M_B < M_{B,cr}$ is the same as the corresponding neutrino-free PHS with equal baryonic mass. 

\begin{figure}
\vspace{0.5cm}
\resizebox{0.43\textwidth}{!}{%
\includegraphics{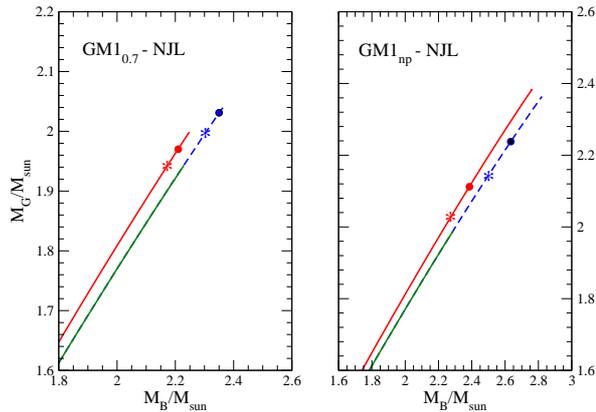} }
\vspace{0.2cm}
\caption{(Color on line) Same as in the previous figure but in the case of the NJL model for 
the quark phase and the Glendenning--Moszkowski model for hyperonic matter with $x_{\sigma}=0.7$ 
(GM1$_{0.7}$) (left panel) and pure nucleonic matter  GM1$_{\rm{np}}$ (right panel).} 
\label{Mg_Mb_GM1_NJL} 
\end{figure}

In Fig.\ \ref{Mg_Mb_GM1_NJL} we plot the PHS, cold HS, and cold QS sequences in the 
gravitational--baryonic mass plane for the case of the NJL model for the quark phase and 
the GM1 model in the case of pure nucleonic matter (right panel) or hyperonic matter 
with $x_{\sigma}=0.7$ (left panel).  It is clearly seen that in the case of the NJL model 
it is almost impossible to populate the QS branch. Cold quark stars can be formed  
in the case of $x_{\sigma}=0.7$ (left panel) for a very narrow range of baryonic stellar 
masses $ 2.20 <  M_B/M_{sun} < 2.23\,$.  

In summary, proto-hadronic stars with a gravitational mass lower than the critical mass $M_{cr}$ could 
survive the early stages of their evolution without decaying to a quark star \cite{bom+09,bom+11}.   
This outcome contrasts with the predictions of the earlier studies \cite{ho92,ho94,ol94,harko04} 
where it was inferred that all the pure hadronic compact stars, with a central temperature above  
2 -- 3 MeV,  are converted to quark stars within the first seconds after their birth. 
However, the prompt formation of a critical size drop of quark matter could occur when 
$M > M_{cr}$. These proto-hadronic stars evolve to cold and deleptonized quark stars  
or collapse to a black holes \cite{bom+09,bom+11}.  

Finally, if quark matter nucleation occurs during the post-bounce stage of core-collapse supernova, 
then the quark deconfinement phase transition could trigger a delayed supernova explosion 
characterized by a peculiar neutrino signal \cite{sage+09,min+10,Naka08,dasg10}.

\section{Conclusions}
In the present review, which mainly summarizes the research reported in  
Ref.~\cite{be02,be03,bo04,vbp05a,vbp05b,lug05,blv07,bppv08,bom+09,bom10,bom+11,log+12a,log+12b,log+13},  
we have investigated the consequences of the quark deconfinement phase transition in stellar compact objects 
when finite size effects between the deconfined quark phase and the hadronic phase are taken into account 
considering a first order phase transition.      

We have found that above a threshold value of the gravitational mass a pure hadronic star is metastable 
to the decay (conversion) to a quark star. 
We have calculated the {\it mean-life time} of these metastable stellar configurations, 
the critical mass for the hadronic star sequence, and have explored how these quantities  
depend on the details of the EOS for dense matter.  
We have exposed how to extend \cite{bo04} the concept of limiting mass of compact stars,   
with respect to the classical one given by Oppenheimer and Volkoff \cite{ov39}. 

The stellar conversion of a HS to a QS liberates an energy of the order of 10$^{53}$~erg and 
will cause a powerful neutrino burst, likely accompanied by intense gravitational waves emission, 
and conceivably it could cause a second delayed explosion (with respect to the supernova explosion, 
which formed the hadronic star). Under opportune physical conditions this second explosion could be 
the energy source of a powerful gamma-ray burst \cite{be02,be03}. 
This scenario is thus able to explain a "delayed" connection between supernova explosions 
and GRBs \cite{be02,be03}.  
It could also explain \cite{bom-popov04} in a natural way the observed bimodal distribution 
of the kick velocities of radio pulsars \cite{arzou02}. 

In addition, within the scenario discussed in the present review, 
one has two coexisting families of compact stars: 
pure hadronic stars and quark stars \cite{be02,be03,bo04} 
(see also the contribution of Drago et al. in this Topical Issue). 
The members of these two families could have similar values for their gravitational masses 
but different values for their radii.   

 The advanced and innovative instruments on board of the new generation of X-ray satellites, as the 
Astrosat satellite (Indian Space Research Organization, which has been launched on September 28$^{th}$ 2015), 
or on board of the NICER satellite (NASA, launch scheduled for October 2016) 
will be able to determine neutron star radii with high precision ({\it i.e.} better than 1~km uncertainty).  
Thus accurate measurements of both the mass and radius of a few individual "neutron stars" \cite{Bhat2010,SLB10} 
or measurements of temperature profiles of accretion discs around rapidly spinning compact stars \cite{BTB2001} could shed light on the validity of the scenario discussed in the present work.

\section*{Acknowledgments} 
This work is partly supported by NewCompstar, COST Action MP1304. 


\end{document}